\documentclass[12pt]{article}
\usepackage{epsf}
\usepackage{amsmath}
\usepackage{a4wide,graphicx}
\usepackage{cite}
\begin{document}

\title{Couplings in coupled channels versus wave functions in the case of resonances: application to the two $\Lambda(1405)$ states}

\author{J. Yamagata-Sekihara$^{1,2}$, J. Nieves$^2$ and E. Oset$^{1,2}$\\
$^1$Departamento de Fisica Teorica, Universidad de Valencia\\ 
$^2$Instituto de F{\'\i}sica Corpuscular (centro mixto CSIC-UV)\\
Institutos de Investigaci\'on de Paterna, Aptdo. 22085, 46071, Valencia, Spain 
}

\date{\today}

\maketitle

 \begin{abstract}
 In this paper we develop a formalism to evaluate wave functions in momentum and coordinate space for the resonant states dynamically generated in a unitary coupled channel approach. The on shell approach for the scattering matrix, commonly used, is also obtained in Quantum Mechanics with a separable potential, which allows one to write wave functions in a trivial way. We develop useful relationships among the couplings of the dynamically generated resonances to the different channels and the wave functions at the origin. The formalism provides an intuitive picture of the resonances in the coupled channel approach, as bound states of one bound channel, which decays into open ones. It also provides an insight and practical rules for evaluating couplings of the resonances to external sources and how to deal with final state interaction in production processes. As an application of the formalism we evaluate the wave functions of the two $\Lambda(1405)$ states in the $\pi \Sigma$, $\bar{K} N$ and other coupled channels. 
It also offers a practical way to study three body systems when two of them cluster into a resonance.
\end{abstract}

\section{Introduction}

The chiral unitary approach to hadron dynamics has brought a new perspective to deal with the interaction of hadrons and the nature of some resonant mesonic
\cite{npa,ramonet,norbert,markushin,juan2,juanenrique,luisaxial} and baryonic states
 \cite{Kaiser:1995cy,weise,Kaiser:1996js,angels,ollerulf,carmina,carmenjuan,
 hyodo,Hyodo:2006kg}
which appear dynamically generated from the interactions and, thus, have a nature quite different to standard $q \bar{q}$ states. With some different formulations at the beginning \cite{Kaiser:1995cy,weise,juanenrique}, the more recent work uses the on shell formulation firstly established on the basis of the N/D method in \cite{nsd}, where the potential and the t-matrix in momentum space factorize outside the loop function implicit in the Bethe Salpeter equation in coupled channels that one uses in those approaches.  This is a very practical way to deal with the problem since one renders the coupled integral equations into a set of algebraic equations, paying a small price which is the fine tuning of some subtraction constant appearing in the dispersion relations involved. The approach is practical and useful, but carries also a handicap which is that one deals with amplitudes in momentum space, and couplings of the dynamically generated resonances to the different channels, and nowhere do wave functions in coordinate space appear in the approach. For instance, all properties of a resonance are given in terms of the mass and width and its couplings to the different channels, obtained from the residues of the amplitudes at the poles in the complex plane. Any intuition about the wave function of the different channels, its magnitude and extent in space, is lost in the approach as well as the meaning of the resonances and the discrete values obtained for the resonant energies.  Obviously, these are magnitudes that help to understand the microscopical composition of the dynamically generated states and, hence, a most welcome information.
However, this is not all, since the scattering amplitudes do not contain 
all the information of the wave function. They reflect the wave function 
at long distances. For some observables the wave function at small 
distances is needed. This is the case when one studies the response of 
states to external sources, which require to evaluate form factors, or 
expectation values of different observables. The wave functions, whether 
in momentum space or coordinate space, are then needed. To study 
couplings of the resonance to local sources the wave function around the 
origin is required, but in the study of form factors one needs to know 
the wave function at all distances. We provide them both in the present 
work. They are already proving very useful to study new systems with 
three or more particles, when two of them cluster to form one of these 
dynamically generated resonances \cite{multirho}.

The first steps in this direction were done in \cite{conenrique} in order to understand the X(3872) resonance in terms of 
a molecule of $D^0 \bar{D^*}^0$ and $D^+ D^{*-}$ and their charge conjugates. 
The reason is that the $D^0 \bar{D^*}^0$ component is slightly 
bound, while the  $D^+ D^{*-}$ one is bound by about 7 MeV. This 
has as a consequence that the $D^0 \bar{D^*}^0$ component stretches up 
to very large distances, while the $D^+ D^{*-}$ one is more 
confined, and this has important repercussion in the interpretation of 
observables regarding this resonance.

The work of \cite{conenrique} provided also an extension of the two channel ($D^0 \bar{D^*}^0$ and $D^+ D^{*-}$) formalism to a general one with many coupled channels, but all of them bound. Hence, in \cite{conenrique} one deals with bound states which as a consequence have a discrete spectrum of energies, and the wave function was only evaluated for the eigenstates of the system. It is most advisable to extend the approach to the case where one has also coupled channels which are open at a certain energy, which is the majority of the cases in the studies done with the chiral unitary approach. One might think that the problem is formally identical to the one of the bound coupled channels, but, although there are certainly analogies, there are also subtle, and important, differences which call for a detailed study. The first big difference is that now, one has wave functions, those of the open channels, that extend to infinity, while the wave functions of the bound channels will remain constrained in space. The wave functions for the open channels will not be finite normalizable, while those of the bound channels with remain finite normalizable.  This problem has been found before and the difficulties of dealing 
together with bound states and open, unnormalizable states, has been 
pointed out \cite{vanBeveren:1983td,Verschuren:1991bg}. We have found a 
way to deal with this problem and answered questions like:  What are the magnitudes that matter in this case when it comes to evaluate observables in terms of the wave functions? What is the meaning of the resonances in this approach, and why the number of resonant energies are finite? These are some of the novel questions that one may now ask and which have to be properly addressed.  This is the purpose of the present work. We shall develop the formalism and then we will apply the results to evaluate the wave functions in the ten coupled channels of the two $\Lambda(1405)$ states obtained in \cite{carmenjuan,cola}.
  
\section{Formalism}
\label{sec:2}
We follow closely the formalism of \cite{conenrique} adapting it to the case of open channels.
First we will consider the relation between a coupling constant and the wave function in one channel case.
In a second step we will extend the formalism to the case of multiple coupled channels.

\subsection{One channel case}
As in \cite{conenrique}, we take a potential $V$ in $s$-wave as a separable function in momentum space with the modulating factor being a step function,
\begin{equation}
\langle \mbox{\boldmath $p'$}|V|\mbox{\boldmath $p$}\rangle=V({\mbox{\boldmath $p'$}},{\mbox{\boldmath $p$}})=v\theta(\Lambda-p')\theta(\Lambda-p)
\label{eq:1}
\end{equation}
where $p$ and $p'$ indicate the moduli of the three momenta ${\mbox{\boldmath $p$}}$ and ${\mbox{\boldmath $p'$}}$, $\Lambda$ is a cutoff in momentum space and $\theta(\Lambda-p)$ is the step function.
In Eq.~(\ref{eq:1}), $v$ does not depend on momentum.
It is worth stressing that the use of this simple form is sufficient to develop the formalism and it has the advantage that it leads to the same on shell factorized Bethe Salpeter equations (here we shall use their non relativistic Lippmann Schwinger form) than in the chiral unitary approach.

\subsection{The wave function}
The Schr${\ddot{\rm o}}$dinger equation reads
\begin{equation}
(H_0+V)|\psi\rangle=E|\psi\rangle
\label{eq:2}
\end{equation}
where $H_0$ is the kinetic term, $V$ is the potential, $\psi$ the exact wave function of the full Hamiltonian $H=H_0+V$ and $E$ the energy.
Since we have open channels we have solutions for any value of $E$.
In order to derive the Lippmann Schwinger equation we proceed as usual, introducing the solution $\phi$ of the kinetic energy Hamiltonian $H_0$ for the same energy $E$.
\begin{equation}
H_0|\phi\rangle=E|\phi\rangle~~.
\label{eq:3}
\end{equation}
From these equations, we obtain
\begin{eqnarray}
(E-H_0)|\psi-\phi\rangle&=&V|\psi\rangle\\
|\psi-\phi\rangle&=&\frac{1}{E-H_0}V|\psi\rangle\\
|\psi\rangle&=&|\phi\rangle+\frac{1}{E-H_0}V|\psi\rangle
\label{eq:6}
\end{eqnarray}
and the wave function in momentum space,
\begin{eqnarray}
\langle \mbox{\boldmath $p$}|\psi\rangle &=&\langle \mbox{\boldmath $p$}|\phi\rangle+\int d^3p' d^3p''\langle \mbox{\boldmath $p$}|\frac{1}{E-H_0}|\mbox {\boldmath $p$}'\rangle \langle \mbox{\boldmath $p$}'|V|\mbox{\boldmath $p$}''\rangle \langle \mbox{\boldmath $p$}''|\psi\rangle \nonumber\\
&=&\langle \mbox{\boldmath $p$}|\phi\rangle+\frac{\theta(\Lambda-p)}{E-m_1-m_2-\mbox {\boldmath $ p$}^2/2\mu+i\epsilon}v\int_{p''<\Lambda}d^3p''\langle \mbox{\boldmath $p$}''|\psi\rangle
\label{eq:7}
\end{eqnarray}
where $\mu$ is the reduced mass of the two particles.
In the bound state case ($E<m_1+m_2$) the term $\langle \mbox{\boldmath $p$}|\phi\rangle$ does not appear, $E-m_1-m_2-\mbox{\boldmath $p$}^2/2\mu$ cannot be zero for any energy and for some energies one finds discrete eigenstates.
However now, any energy $E>m_1+m_2$ is allowed.
Let us integrate Eq.~(\ref{eq:7}) over $\mbox{\boldmath $p$}$,
\begin{eqnarray}
\int_{p<\Lambda}d^3p\langle \mbox{\boldmath $p$}|\psi \rangle &=&\theta\left(\Lambda-\sqrt{2\mu(E-m_1-m_2)}\right)\nonumber\\
&&+\int_{p<\Lambda}d^3p\frac{1}{E-m_1-m_2-\mbox {\boldmath $\scriptsize p$}^2/2\mu}v\int_{p''<\Lambda}d^3p''\langle \mbox{\boldmath $p$}''|\psi\rangle~\nonumber\\
 &=&\theta\left(\Lambda-\sqrt{2\mu(E-m_1-m_2)}\right)+Gv\int_{p''<\Lambda}d^3p''\langle \mbox{\boldmath $p$}''|\psi\rangle~~,
\label{eq:8}
\end{eqnarray}
where
\begin{equation}
G=\int_{p<\Lambda}d^3p\frac{1}{E-m_1-m_2-\mbox{\boldmath $\scriptsize p$}^2/2\mu+i\epsilon}~~.
\label{eq:G}
\end{equation}
the term $\theta\left(\Lambda-\sqrt{2\mu(E-m_1-m_2)}\right)$ in Eq.~(\ref{eq:8}) comes because of the normalization that we impose on the states
\begin{equation}
\langle \mbox{\boldmath $p$}|\mbox{\boldmath $p'$}\rangle=\delta^3(\mbox{\boldmath $p$}-\mbox{\boldmath $p'$})
\label{eq:10}
\end{equation}
since $|\phi \rangle=|\mbox{\boldmath $p'$}\rangle$ such that $\mbox{\boldmath $p'$}^2/2\mu+m_1+m_2=E$.
Since the integration in Eq.~(\ref{eq:G}) has the singularity at $E=m_1+m_2+\mbox{\boldmath $\scriptsize p$}^2/2\mu$, we put $+i\epsilon$ which guarantees an outgoing solution for the Lippmann Schwinger equation.
In what follows, we will assume $\Lambda>\sqrt{2\mu(E-m_1-m_2)}$.
This is essential to get the unitarity properties from the $G$ function of Eq.~(\ref{eq:G}) and we must be certain that this occurs for all channels in the coupled channel case.
From Eq.~(\ref{eq:8}) we have now
\begin{eqnarray}
(1-Gv)\int_{p<\Lambda}d^3p\langle\mbox{\boldmath $p$}|\psi\rangle&=&1\\
\int_{p<\Lambda}d^3p\langle\mbox{\boldmath $p$}|\psi\rangle&=&\frac{1}{1-Gv}
\label{eq:12}
\end{eqnarray}
and if $G$ is complex this means that $\int_{p<\Lambda}d^3p \langle \mbox{\boldmath ${p}$}|\psi \rangle$ is complex and hence $\langle \mbox{\boldmath${p}$}|\psi\rangle$ is necessarily complex.

From Eq.~(\ref{eq:7}), the wave function in coordinate space can be equally evaluated by
\begin{eqnarray}
\langle \mbox{\boldmath $x$}|\psi\rangle&=&\int d^3p \langle \mbox{\boldmath $x$}|\mbox{\boldmath $p$}\rangle\langle \mbox{\boldmath $p$}|\psi\rangle \nonumber\\
&=&\frac{1}{(2\pi)^{3/2}}e^{i \mbox{\boldmath \scriptsize$p'$}\cdot\mbox{\boldmath \scriptsize$x$}}\nonumber\\
&&+\int_{p<\Lambda}d^3p\frac{1}{(2\pi)^{3/2}} e^{i\mbox{\boldmath \scriptsize$p$}\cdot \mbox{\boldmath \scriptsize$x$}}\frac{1}{E-m_1-m_2-\mbox{\boldmath$p$}^2/2\mu+i\epsilon}v\frac{1}{1-Gv}
\label{eq:13}
\end{eqnarray}
and we can obtain the value of the wave function at the origin in coordinate space
\begin{equation}
\langle \mbox{\boldmath $0$}|\psi\rangle=\frac{1}{(2\pi)^{3/2}}+\frac{1}{(2\pi)^{3/2}}Gv\frac{1}{1-Gv}~~.
\end{equation}

Now we define $\hat \psi=(2\pi)^{3/2}\psi(\mbox{\boldmath $0$})$ and we obtain
\begin{equation}
\hat \psi=(2\pi)^{3/2}\psi(\mbox{\boldmath $0$})=\frac{1}{1-Gv}~~.
\label{eq:15}
\end{equation}
and by means of Eq.~(\ref{eq:12})
\begin{equation}
\hat\psi=\int_{p<\Lambda}d^3p\langle\mbox{\boldmath $p$}|\psi\rangle
\label{eq:a}
\end{equation}
Eq.~(\ref{eq:a}) is also found in the case of bound states, but Eq.~(\ref{eq:15}) is new for scattering.
For the case of a bound state one finds in \cite{conenrique} $\hat\psi=gG$, where $g$ is the coupling of the bound state to the channel considered.
This difference must be clarified.
In Eq.~(\ref{eq:3}), we considered a plane wave $\phi$, which is needed to satisfy the boundary condition at infinity.
For the bound wave function, $\phi$ does not appear because $(E-H_0)|\phi\rangle =0$ has no solution for $E<m_1+m_2$. 
In this case the Schr$\ddot{\rm o}$dinger equation has only a few discrete eigenenergies $E_\alpha$.

\subsubsection{The coupling}
Let us consider the Lippmann Schwinger equation.
Let us define $T$ such that $T|\phi\rangle=V|\psi\rangle$.
From Eq. (\ref{eq:6}), we obtain
\begin{eqnarray}
T|\phi\rangle&=&V|\phi\rangle+V\frac{1}{E-H_0}V|\psi\rangle\nonumber\\
T&=&V+V\frac{1}{E-H_0}T~~.
\label{eq:16}
\label{eq:Lip}
\end{eqnarray}
From Eq.~(\ref{eq:Lip}), we can write 
\begin{equation}
\langle \mbox{\boldmath $p$}|T|\mbox{\boldmath $p'$}\rangle=\langle \mbox{\boldmath $p$}|V|\mbox{\boldmath $p'$}\rangle+\int d^3p''\frac{\langle \mbox{\boldmath $p$}|V|\mbox{\boldmath $p''$}\rangle}{E-m_1-m_2-\mbox{\boldmath $p''$}^2/2\mu}\langle \mbox{\boldmath $p''$}|T|\mbox{\boldmath $p'$}\rangle
\end{equation}
which has the solution 
\begin{equation}
\langle \mbox{\boldmath $p$}|T|\mbox{\boldmath $p'$}\rangle \equiv \theta(\Lambda-p')\theta(\Lambda-p)t~~,
\label{eq:18}
\end{equation}
and then
\begin{equation}
t=v+vGt=\frac{v}{1-vG}=\frac{1}{v^{-1}-G}~~.
\label{eq:t}
\end{equation}
For real $E$ we do not have poles since $vG$ is complex for real $E$.
However, when we have $1-{\rm Re~}vG\simeq 0$ we have a near pole and hence an enhancement of $t$, and thus a resonance.
As we know, we can even find a pole in the complex plane in the second Riemann sheet associated to this resonance.

In the vicinity of the pole, $t$ also can be written as
\begin{equation}
t=\frac{g^2}{E-E_{\rm R}+i\Gamma/2}\equiv \frac{1}{v^{-1}-G}~~.
\label{eq:t_pole}
\end{equation}
We define the resonance energy $E_{\rm R}$ when one satisfies $v^{-1}-{\rm Re}~G(E_{\rm R})=0$. 
Using the equation below, and assuming $v$ constant as a function of energy
\begin{equation}
v^{-1}-{\rm Re}~G(E)\simeq v^{-1}-{\rm Re}~G(E_{\rm R})-\left.\frac{\partial {\rm Re}~G}{\partial E}\right|_{E_{\rm R}}(E-E_{\rm R})+\cdots~~,
\end{equation}
we rewrite Eq.~(\ref{eq:t_pole}) as
\begin{equation}
t\simeq\frac{1}{-\frac{\partial {\rm Re}~G}{\partial E}|_{E_{\rm R}}(E-E_{\rm R})-i{\rm Im}~G(E_{\rm R})}
\end{equation}
which gives us the two conditions
\begin{eqnarray}
g^2&=&-\Bigl(\frac{\partial {\rm Re}~G}{\partial E}\Bigr)_{E_{\rm R}}^{-1}\\
\frac{\Gamma}{2}&=&-g^2{\rm Im} G~~.
\end{eqnarray}

The integral for the $G$ function defined in Eq.~(\ref{eq:G}) can be performed analytically and we obtain
\begin{eqnarray}
G(E)&=&-8\pi\mu(\Lambda+\frac{k}{2}\ln\frac{\Lambda-k}{\Lambda+k})-i4\pi^2\mu k
\label{eq:26}\\
k&=&\sqrt{2\mu (E-m_1-m_2)}~.
\end{eqnarray}
It is instructive to check that an analytical extrapolation below threshold of Eq.~(\ref{eq:26}), putting $k=\pm i\gamma$ ($\gamma=\sqrt{2\mu B}$, with $B$ the binding energy), leads to the formula for the $G$ function for bound states (Eq.~(27) of \cite{conenrique}).
The above equation gives us the coupling $g$ as
\begin{equation}
g^2=\frac{k_{\rm R}}{8\mu^2\pi\left[\ln\Bigl((\Lambda-k_{\rm R})/(\Lambda+k_{\rm R})\Bigr)/2-k_{\rm R}\Lambda/(\Lambda^2-k_{\rm R}^2)\right]}~~.
\end{equation}
The first thing we observe is that $g^2<0$ and $\Gamma<0$, which is not physical.
The result obtained simply expresses the fact that a constant potential $v$ as a function of energy does not lead to a resonance (an energy dependent $v$ could give a resonance).
However, with $v$ negative one can obtain bound states as found in~\cite{conenrique}.
A resonance state in one channel is usually associated to a barrier in coordinate space which is not reproduced by $v$ constant.
The situation will be very different when we go to coupled channels.

The other feature we must to mention is that the limit of $g^2$ when $k_{\rm R}\to 0$ is given by
\begin{equation}
g^2_{k_{\rm R}\to 0}=-\frac{\Lambda}{16\mu^2\pi}
\end{equation}
which is not zero, unlike the case of bound states~\cite{conenrique}, or the case that we shall see in coupled channels, where the existence of only one bound channel guarantees $g^2\to 0$ for all channels including those with energies above the threshold of the bound channel.

We can also obtain the asymptotic behavior of the wave function at $r\to \infty$ from Eq.~(\ref{eq:13}) and we find
\begin{equation}
(2\pi)^{3/2}\langle\mbox{\boldmath $x$}|\psi\rangle \xrightarrow[r\to \infty]{}e^{i \mbox{\boldmath \scriptsize$p'$}\cdot\mbox{\boldmath \scriptsize$x$}}-4\pi^2\mu\frac{v}{1-vG}\frac{e^{ip'r}}{r}
\end{equation}
which tells us that the scattering matrix $f(\theta)$ of Quantum Mechanics is given by
\begin{equation}
f(\theta)=-4\pi^2\mu\frac{v}{1-vG}=-4\pi^2\mu t~~.
\end{equation}

\subsection{Coupled Channels}
We extend now the formalism to $N$ coupled channels, where at least one of them is bound.
We take again
\begin{equation}
\langle \mbox{\boldmath $p'$}|V|\mbox{\boldmath $p$}\rangle=\theta(\Lambda-p)\theta(\Lambda-p')v
\label{eq:v32}
\end{equation}
where $v$ is now a $N\times N$ matrix.
\subsubsection{The wave function}
The first thing when dealing with coupled channels is that one must find the boundary conditions for the physical process that one is studying.
If we wish to create a resonance from the interaction of many channels at a certain energy we must take a channel which is open at this energy and make the two particles collide, starting from an infinite separation at $t=-\infty$.
Let us call channel 1 to this open channel that undergoes the scattering.
The equations to solve, with the appropriate boundary condition of a scattering state for channel 1 are
\begin{equation}
|\psi\rangle=|\phi\rangle+\frac{1}{E-H_0}V|\psi\rangle
\end{equation}
where
\begin{equation}
|\psi\rangle\equiv
\begin{Bmatrix}
|\psi_1\rangle\\
|\psi_2\rangle\\
\vdots \\
|\psi_N\rangle\\
\end{Bmatrix}
~~,|\phi\rangle\equiv
\begin{Bmatrix}
|\phi_1\rangle\\
0\\
\vdots \\
0\\
\end{Bmatrix}
\end{equation}
and $|\phi_1\rangle=|\mbox{\boldmath $p'$}\rangle$, such that $\mbox{\boldmath $p'$}^2/2\mu_1+M_1=E$, where we will use the notation $M_i=m_{1i}+m_{2i}$ and $\mu_i=m_{1i}m_{2i}/(m_{1i}+m_{2i})$.

Following Eq.~(\ref{eq:7}), the wave functions in momentum space are written
\begin{eqnarray}
\langle \mbox{\boldmath $p$}|\psi_1\rangle&=&\langle \mbox{\boldmath $p$}|\phi_1\rangle+\frac{\theta(\Lambda-p)}{E-M_1-\mbox {\boldmath $ p$}^2/2\mu_1+i\epsilon}\sum_jv_{1j}\int_{p''<\Lambda}d^3p''\langle \mbox{\boldmath $p$}''|\psi_j\rangle
\label{eq:31}\\
\langle \mbox{\boldmath $p$}|\psi_i\rangle&=&\frac{\theta(\Lambda-p)}{E-M_i-\mbox {\boldmath $ p$}^2/2\mu_i+i\epsilon}\sum_jv_{ij}\int_{p''<\Lambda}d^3p''\langle \mbox{\boldmath $p$}''|\psi_j\rangle~~(i\ne 1)~.
\label{eq:32}
\end{eqnarray}
And integrating over $\mbox{\boldmath $p$}$, we have
\begin{eqnarray}
\int_{p<\Lambda} d^3p\langle \mbox{\boldmath $p$}|\psi_1\rangle&=&1+G_{11}\sum_jv_{1j}\int_{p''<\Lambda}d^3p''\langle \mbox{\boldmath $p$}''|\psi_j\rangle
\label{eq:33}\\
\int_{p<\Lambda} d^3p\langle \mbox{\boldmath $p$}|\psi_i\rangle&=&G_{ii}\sum_jv_{ij}\int_{p''<\Lambda}d^3p''\langle \mbox{\boldmath $p$}''|\psi_j\rangle~~(i\ne 1)~~,
\label{eq:34}
\end{eqnarray}
where once again we assume $\Lambda$ to be bigger that the on shell momenta of the particles for all the open channels.
In Eqs.~(\ref{eq:33}), (\ref{eq:34}) the diagonal $G$ matrix is given by
\begin{equation}
G= \bordermatrix{
&&&&\cr
&G_1&&&\cr
&&G_2&&\cr
&&&\ddots&\cr
&&&&G_N\cr
}~~.
\end{equation}
with $G_i$ given by Eq.~(\ref{eq:G}) for each channel.
Thus, we have
\begin{eqnarray}
\hat{\psi _1}&=&1+G_{11}\sum_jv_{1j}\hat{\psi _j}\\
\hat{\psi _i}&=&G_{ii}\sum_jv_{ij}\hat{\psi _j}~~(i\ne 1)
\label{eq:coupled_psi}
\end{eqnarray}
where $\hat \psi_i=\int_{p<\Lambda}d^3p\langle \mbox{\boldmath $p$}|\psi_i\rangle$ and Eq.~(\ref{eq:coupled_psi}) is then written in matrix form as
\begin{equation}
\{ \hat \psi \}=(1-Gv)^{-1}
\begin{Bmatrix}
1\\
0\\
\vdots \\
0\\
\end{Bmatrix}
\end{equation}
which provides the values of $\hat\psi_i$ as
\begin{equation}
\hat\psi_i=(1-Gv)^{-1}_{i1}~~.
\label{eq:39}
\end{equation}
The scattering matrix in coupled channels is given formally by Eq.~(\ref{eq:16}), but $T$ is now a $N\times N$ matrix.
It has also the form of Eq.~(\ref{eq:18}) and now the $N\times N$ $t$ matrix is given by
\begin{equation}
t=[1-vG]^{-1}v=[v^{-1}-G]^{-1}~~.
\end{equation}
By means of this equation, the combination $v\hat\psi$ appearing in Eqs.~(\ref{eq:31}), (\ref{eq:32}) can be written as
\begin{eqnarray}
v_{ij}\hat\psi_j&=&v_{ij}(1-Gv)^{-1}_{j1}=[v(1-Gv)^{-1}]_{i1}\nonumber\\
&=&[vv^{-1}(v^{-1}-G)^{-1}]_{i1}=[(v^{-1}-G)^{-1}]_{i1}=t_{i1}
\end{eqnarray}

Taking this into account, the wave function equations of Eqs.~(\ref{eq:31}), (\ref{eq:32}) become
\begin{eqnarray}
\langle \mbox{\boldmath $p$}|\psi_1\rangle&=&\langle \mbox{\boldmath $p$}|\phi_1\rangle+\frac{\theta(\Lambda-p)}{E-M_1-\mbox{\boldmath $p$}^2/2\mu_1+i\epsilon}t_{11}(E)\\
\langle \mbox{\boldmath $p$}|\psi_i\rangle&=&\frac{\theta(\Lambda-p)}{E-M_i-\mbox{\boldmath $p$}^2/2\mu_i+i\epsilon}t_{i1}(E)~~(i\ne 1)
\end{eqnarray}
and we can write the wave function in coordinate space as

\begin{eqnarray}
\langle \mbox{\boldmath $x$}|\psi_1\rangle&=&\frac{1}{(2\pi)^{3/2}}e^{i \mbox{\boldmath \scriptsize$p'$}\cdot\mbox{\boldmath \scriptsize$x$}}+\int_{p<\Lambda}d^3p\frac{1}{(2\pi)^{3/2}}e^{i \mbox{\boldmath \scriptsize$p$}\cdot\mbox{\boldmath \scriptsize$x$}}\frac{1}{E-M_1-\mbox{\boldmath $p$}^2/2\mu_1+i\epsilon}t_{11}(E)
\label{eq:xWF1}\\
\langle \mbox{\boldmath $x$}|\psi_i\rangle&=&\int_{p<\Lambda}d^3p\frac{1}{(2\pi)^{3/2}}e^{i \mbox{\boldmath \scriptsize$p$}\cdot\mbox{\boldmath \scriptsize$x$}}\frac{1}{E-M_i-\mbox{\boldmath $p$}^2/2\mu_i+i\epsilon}t_{i1}(E)~~(i\ne 1)
\label{eq:xWFi}
\end{eqnarray}
with $|\mbox{\boldmath $p'$}|=\sqrt{2\mu_1(E-M_1)}$.

Once again we can make the limit $\mbox{\boldmath $r$}\to\infty$ and we find the asymptotic solutions
\begin{eqnarray}
(2\pi)^{3/2}\langle\mbox{\boldmath $x$}|\psi_1\rangle&\xrightarrow[r\to\infty]{}&e^{i \mbox{\boldmath \scriptsize$p'$}\cdot\mbox{\boldmath \scriptsize$x$}}-4\pi^2\mu_1\frac{e^{ip'r}}{r}t_{11}\\
(2\pi)^{3/2}\langle\mbox{\boldmath $x$}|\psi_i\rangle&\xrightarrow[r\to\infty]{}&-4\pi^2\mu_i\frac{e^{ik_ir}}{r}t_{i1}~~{\rm (open~channel)}
\label{eq:c}\\
&&-4\pi^2\mu_i\frac{e^{-\kappa_ir}}{r}t_{i1}~~{\rm (bound~state)}\nonumber
\end{eqnarray}
where $k_i=\sqrt{2\mu_i(E-M_i)}$ for open channels and $\kappa_i=\sqrt{2\mu_i|E-M_i|}$ for the bound channels.

As we can see, the channel 1 contains the plane wave from the scattering state while the other channels only have the wave functions generated from the collision of the particles of the scattering channel.
For bound channels one generates a bound wave function while in the other channels we have an outgoing wave.
Eqs.~(\ref{eq:c}) show the connection of the scattering matrix $t$ with that of Quantum Mechanics $f_{i1}$, such that
\begin{equation}
(2\pi)^{3/2}\langle \mbox{\boldmath $x$}|\psi_i\rangle \xrightarrow[r\to \infty]{}\sqrt{\frac{\mu_i}{\mu_1}}f_{i1}(\theta)\frac{e^{i k_ir}}{r}
\label{eq:a1}
\end{equation}
which implies
\begin{equation}
\left.\frac{d\sigma}{d\Omega}\right|_{1\to i}=\frac{k_i}{k_1}\left|f_{i1}(\theta)\right|^2~.
\label{eq:a2}
\end{equation}
Thus we have 
\begin{equation}
f_{i1}(\theta)=-4\pi^2\sqrt{\mu_i\mu_1}t_{i1}
\label{eq:a3}
\end{equation}
with no angle dependence since we are dealing with $s$-waves from the beginning.

We can go back to Eqs.~(\ref{eq:xWF1}), (\ref{eq:xWFi}), by taking $(2\pi)^{3/2}\langle \mbox{\boldmath $x$}=\mbox{\boldmath $0$}|\psi_i\rangle$, which is $\hat{\psi_i}$ (see Eqs.~(\ref{eq:15}), (\ref{eq:a})), and calling $\hat{\psi}_1^{\rm (out)}$ the outgoing wave of $\langle \mbox{\boldmath $x$}|\psi_1\rangle$ at $\mbox{\boldmath $x$}=0$ (term of $\int d^3\mbox{\boldmath $p$}$ in Eq.~(\ref{eq:xWF1})), we obtain
\begin{eqnarray}
\hat{\psi}_1^{\rm (out)}(E)&=&G_1(E)t_{11}(E)
\label{eq:55}\\
\hat{\psi}_i(E)&=&G_i(E)t_{i1}(E)
\label{eq:56}
\end{eqnarray}
which generalizes similar equations obtained for the bound case in \cite{conenrique}, but substituting $t_{11}$, $t_{i1}$ by $g_1$, $g_i$ respectively.
Here $E$ is a continuous variable and we see that the wave functions depend on the energy, but for resonant energies the wave functions at the origin grow roughly like $t$, since $G(E)$ is smoother than $t$ at the resonant peak.
Thus, we get an intuitive idea about the meaning of a resonance, which is that for resonant energies there is an accumulation of strength of the wave functions at the origin.
We can also see that
\begin{equation}
\frac{G_i^{-1}\hat\psi_i}{G_1^{-1}\hat\psi_1^{\rm (out)}}=\frac{t_{i1}}{t_{ii}}\simeq \frac{g_i}{g_1}
\end{equation}
the last part of the equation holding if we can represent the $t_{i1}$ amplitude approximately as $g_ig_1/(\sqrt{s}-M_R+i\Gamma/2)$.

We based all our approach on a potential with a sharp cut off of Eqs.~(\ref{eq:1}) and (\ref{eq:v32}).
The work can be easily generalized to the use of a potential of the type
\begin{equation}
\langle \mbox{\boldmath $p$}'|V|\mbox{\boldmath $p$}\rangle=vf(\mbox{\boldmath $p$})f(\mbox{\boldmath $p$}')
\label{eq:vtype}
\end{equation}
as also done in \cite{conenrique}.
The modifications are minimal and can be followed from Ref.~\cite{conenrique}, section VII.
The most important for our discussion is that the function $\hat\psi$ that represented the wave function at the origin is replaced by
\begin{eqnarray}
\hat\psi_i^{(f)}&=&\int d^3\mbox{\boldmath $k$}f(\mbox{\boldmath $k$})\langle\mbox{\boldmath $k$}|\psi_i\rangle \nonumber\\
&=&\int d^3\mbox{\boldmath $x$}\psi_i(\mbox{\boldmath $x$})\hat{ f} (\mbox{\boldmath $x$})
\end{eqnarray}
where $\hat{f}(\mbox{\boldmath $x$})$ is the Fourier Transform of $f(\mbox{\boldmath $k$})$.
Thus, the wave function at the origin in the present approach is replaced by the folding of the wave function with the Fourier Transform of the factor $f(\mbox{\boldmath $p$})$, in practice an average of the wave function close to the origin.

\subsection{Width of a resonance and partial decay widths}
Within the chiral unitary approach, a different convention for the $T$ matrix is commonly used.
The relationship between these matrices was given in \cite{conenrique}.
Since many relationships are familiar to practitioners of the chiral theoretical approach in the field theoretical notation, we adopt in this section the latter notation and one has for the case of meson baryon states
\begin{eqnarray}
T_{i1}^{\rm FT}&=&32\pi^3\sqrt{\mu_i\mu_1}\sqrt{s}(2M_{\rm B1}2M_{{\rm B}i})^{-1/2}t_{i1}\nonumber\\
&=&-8\pi\sqrt{s}(2M_{\rm B1}2M_{{\rm B}i})^{-1/2}f_{i1}(\theta)
\label{eq:d}
\end{eqnarray}
where $M_{{\rm B}i}$ are the masses of the baryons.

The optical theorem in the scattering of channel 1 is stated as ($k_1\equiv p'$)
\begin{eqnarray}
{\rm Im~}T^{\rm FT}_{11}&=&-\frac{2k_1\sqrt{s}}{2M_{\rm B1}}\sigma_{\rm tot}\nonumber\\
&=&-\frac{k_1\sqrt{s}}{M_{B1}}\sum^{N'}_{i=1}4\pi|f_{i1}(\theta)|^2\frac{k_i}{k_1}
\end{eqnarray}
where we have made use of Eq.~(\ref{eq:a2}), the $4\pi$ comes from the $d\Omega$ integration and $N'$ are the numbers of open channels.
By using Eqs.~(\ref{eq:a3}), (\ref{eq:d}) we find
\begin{equation}
{\rm Im~}T^{\rm FT}_{11}=-\frac{1}{4\pi}\sum^{N'}_{i=1}\frac{M_{{\rm B}i}}{\sqrt{s}}|T^{\rm FT}_{1i}|^2k_i~~.
\label{eq:g}
\end{equation}

Now assume we have a resonance in the coupled channel problem and hence close to $\sqrt{s}=M_{\rm R}$ we have
\begin{equation}
T^{\rm FT}_{11}=\frac{\tilde{g_1}^2}{\sqrt{s}-M_{\rm R}+i\Gamma/2}~~;~~{\rm Im~}T^{\rm FT}_{11}\Bigr|_{\sqrt{s}=M_{\rm R}}=-\frac{\tilde{g_1}^2}{\Gamma/2}~~.
\end{equation}
On the other hand,
\begin{equation}
T_{1i}^{\rm FT}=\frac{\tilde{g_1}\tilde{g_i}}{\sqrt{s}-M_{\rm R}+i\Gamma/2}~~;~~|T^{\rm FT}_{1i}|^2\Bigr|_{\sqrt{s}=M_{\rm R}}=-\frac{\tilde{g_1}^2\tilde{g_i}^2}{(\Gamma/2)^2}
\end{equation}
with $\tilde g_i$ the coupling of the resonance to channel $i$ in the field theoretical convention.

By means of these equations we find
\begin{equation}
\frac{\Gamma}{2}=\sum^{N'}_{i=1}\frac{1}{4\pi}\frac{M_{{\rm B}i}}{M_{\rm R}}\tilde{g_i}^2k_i=\sum^{N'}_{i=1}\frac{\Gamma_i}{2}
\end{equation}
the last equation holding since in this notation the partial decay width of the resonance to an open channel is given by
\begin{equation}
\Gamma_i=\frac{1}{2\pi}\frac{M_{{\rm B}i}}{M_{\rm R}}\tilde{g_i}^2k_i~~.
\label{eq:59}
\end{equation}

\subsection{Coupling of a resonance to external channels and final state interaction considerations}
\begin{figure}[htpd]
\begin{center}
\includegraphics[width=12.0cm,height=1.5cm]{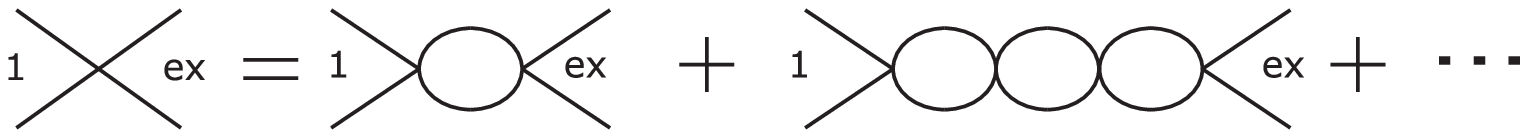}
\caption{\label{fig:m}}
\end{center}
\end{figure}

Sometimes one wishes to evaluate the coupling of the resonance to an external channel, where the coupling is sufficiently weak not to deserve to be taken into account as one of the coupled channels in the approach.
Since the resonance has been created by coupled channels, the coupling of this external channel to the resonance will come from its couplings to the different channels.
We show the simple case where the external channel couples to the building channels through a zero range interaction and let $t_{i{\rm ex}}$ be the transition matrix from any channel to the external one.
The coupling of channel 1 to the external channel can be depicted by the series of terms of Fig.~\ref{fig:m}.
We can ignore the first term of the series if we are only concerned about the resonance contribution.
Hence we get
\begin{equation}
t_{1{\rm ex}}=\sum^N_{i=1}t_{1i}G_it_{i{\rm ex}}\equiv \sum^N_{i=1}\frac{g_1g_i}{\sqrt{s}-M_{\rm R}+i\Gamma/2}G_it_{i{\rm ex}}
\end{equation}
which we would like to equate to
\begin{equation}
t_{1{\rm ex}}=\frac{g_1g_{\rm ex}}{\sqrt{s}-M_{\rm R}+i\Gamma/2}
\end{equation}
where $g_{\rm ex}$ is the coupling of the resonance to this external channel.
We readily obtain
\begin{equation}
g_{\rm ex}=\sum^N_{i=1}g_iG_it_{i{\rm ex}}~~.
\label{eq:62}
\end{equation}
Note that we have assumed $t_{i{\rm ex}}$ to be of zero range, hence constant in momentum space.
Since in the series of Fig.~\ref{fig:m} we have a $v\theta(q-\Lambda)\theta(q'-\Lambda)$ in each four leg vertex, except for the last one, one guarantees that in the loop function one is implementing the cut off $\Lambda$, including the last loop.
This is interesting to note because in Field Theory, in principle the last loop can be regularized in a different way.
However, in the Quantum Mechanical approach we see that all loops are regularized with the same cut off and the last loop is the same one appearing in the scattering problem, provided the range of $t_{i{\rm ex}}$ is shorter than $1/\Lambda$.
The result of Eq.~(\ref{eq:62}) is often used in problems using the chiral unitary approach \cite{geng} and here we find a justification for it.

\begin{figure}[htpd]
\begin{center}
\includegraphics[width=10.0cm,height=1.5cm]{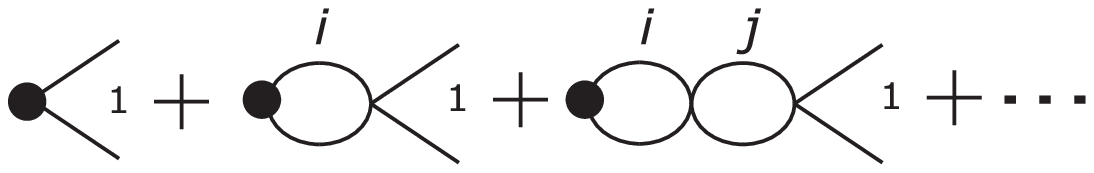}
\caption{\label{fig:h}}
\end{center}
\end{figure}

The other subject worth discussing from the present perspective is the final state interaction.
Assume we have a physical process in which one state is produced.
We assume that the production, for instance in a weak process, is of zero range.
Let us then assume that this state couples to $N$ different channels through strong interaction.
On top of the direct production of the original state,
say channel 1, one could have as well the direct production of any of the coupled channels, which then will make a strong interaction transition to channel 1.
This is depicted in Fig.~\ref{fig:h}.

The full production amplitude will be given by
\begin{equation}
\tilde{P_1}=P_1+\sum^N_{i=1}P_iG_it_{i1}~~.
\label{eq:m}
\end{equation}
Since $P_i$ is considered of zero range, and hence a constant in momentum space, $G_i$ appearing in Eq.~(\ref{eq:m}) is regularized with the same cut off $\Lambda$ in Eq.~(\ref{eq:m}) than in the scattering problem.
This is because all the four leg vertices appearing in Fig.~\ref{fig:h} contain $v\theta(\Lambda-q)\theta(\Lambda-q')$, which impose the $\Lambda$ cut on the loop function.
In principle, in Field Theoretical studies the first loop to the left in Fig~\ref{fig:h} could be regularized in a different way than in the scattering problem.
Yet, the Quantum Mechanical treatment shown here reveals that one can use the same loop function as in the scattering problem provided the primary production vertex corresponds to a short range process, shorter than that implied by the cut off $\Lambda$ used in the scattering.
Eq.~(\ref{eq:m}) is often used in calculations using the chiral unitary approach \cite{gengzou}.

In particular, if we have only one channel, Eq.~(\ref{eq:m}) gives
\begin{equation}
\tilde{P_1}=P_1(1+G_1t_{11})=P_1\frac{t_{11}}{v_{11}}~~,
\end{equation}
which is consistent with Watson's theorem~\cite{WT} and it was also discussed in Ref.~\cite{juanenrique}.
This is the most popular way to implement final state interaction, simply multiply with $t_{11}$ the primary production amplitude, normalizing data at one energy, thus, using only the energy dependence of $t_{11}$ in the analysis \cite{nazo}.
Once again, the present approach shows the conditions upon which such procedure can be trusted; $i)$ Dominance of just one channel, $ii)$ short range nature of the primary production, $iii)$ very weak energy dependence of $v$ with respect to $t$.
This lather condition is flagrantly violated in the case of $\pi\pi$ production around the $\sigma(600)$ region, where $v$ depends more strongly on the energy than $t$, in which case, keeping $v_{11}$ in the $t_{11}/v_{11}$ factor is essential.
This was shown in \cite{jpsione,Li} and it was essential to interpret the apparently narrow $\pi\pi$~$``\sigma"$ structure in the $J/\psi\to\omega\pi^+\pi^-$ production \cite{wu} in terms of the generally admitted wide $\sigma$ resonance \cite{npa,colangelo}.

\subsubsection{The couplings for resonant states}
The scattering matrix is given by the obvious generalization of Eq.~(\ref{eq:t}) for coupled channels and it can be rewritten as
\begin{eqnarray}
t=\frac{Av}{{\rm det}(1-vG)}
\end{eqnarray}
with $A=[{\rm det}(1-vG)](1-vG)^{-1}$.
Now there are open channels and any energy $E$ is allowed.
$E$ is a continuous variable.
Technically we can write
\begin{equation}
T=\frac{Av}{{\rm det}(1-vG)}=\frac{Av}{{\rm Re[det}(1-vG)]+i{\rm Im[det}(1-vG)]}
\end{equation}
and look for $E_{\rm R}$ such that ${\rm Re[det}(1-vG(E_{\rm R}))]=0$.
Around the resonance energy $E_{\rm R}$, we can write
\begin{equation}
{\rm Re[det}(1-vG(E))]={\rm Re[det}(1-vG(E_{\rm R}))]+\frac{\partial}{\partial E}{\rm Re[det}(1-vG(E))]\Bigl|_{E=E_{\rm R}}(E-E_{\rm R})+\cdots
\end{equation}
and then
\begin{equation}
t\simeq\frac{Av}{\frac{\partial}{\partial E}{\rm Re[det}(1-vG(E))]\Bigr|_{E=E_{\rm R}}(E-E_{\rm R})+i{\rm Im[det}(1-vG(E_{\rm R}))]}
\label{eq:Tcoupled1}
\end{equation}
In the vicinity of the pole, $t$ can be written as
\begin{equation}
t_{ij}\simeq\frac{g_ig_j}{E-E_{\rm R}+i\Gamma/2}
\label{eq:Tcoupled2}
\end{equation}
and from Eqs.~(\ref{eq:Tcoupled1}) and (\ref{eq:Tcoupled2}) we obtain the couplings and width
\begin{eqnarray}
g_ig_j&=&\frac{(Av)_{ij}}{\frac{\partial}{\partial E}{\rm Re[det}(1-vG(E))]\Bigr|_{E=E_{\rm R}}}
\label{eq:70}\\
\frac{g_i}{g_j}&=&\frac{(Av)_{ij}}{(Av)_{jj}}\Bigr|_{E=E_{\rm R}}~~,\label{eq:76}\\
\frac{\Gamma}{2}&=&\frac{{\rm Im[det}(1-vG(E))]}{\frac{\partial}{\partial E}{\rm Re[det}(1-vG(E))]}\Bigr|_{E=E_{\rm R}}~~.
\label{eq:50}
\end{eqnarray}

Let us note that some equations relating the couplings obtained for bound states in \cite{conenrique} were a consequence of the dominance of just one eigenstate for the discrete eigenenergies in a sum over intermediate states.
This does not hold here, where $E$ is a continuous variable.
Also other properties where based upon imposition that the total wave function was normalized to unity, which is no longer possible here.
Yet, there is one interesting property which still remains.
Assume we have a resonance state close to the threshold of a bound channel, $a$.
In this case $dG_a/dE\to-\infty$, then $-vdG_a/dE$ dominates in the denominator of Eq.~(\ref{eq:70}) and all the couplings go to zero, as was also the case in the bound states~\cite{conenrique}, and in agreement with the claims made in \cite{toki}.

One can also see that under the dominance of one channel where $|v_{aa}|\gg|v_{ij}|$ ($i,~j\ne a$) and $v_{aa}$ negative, one can find a resonance energy below the threshold $M_a$, and if this is close to threshold, such that $(dG_a/dE)|_{E=E_{\rm R}}$ dominates over all the other terms, one finds approximately,
\begin{equation}
g_a^2\simeq -\frac{1}{\frac{d}{dE}G_a(E)\Bigr|_{E=E_{\rm R}}}
\label{eq:ese}
\end{equation}
like in the case of one channel bound state \cite{conenrique}.
We will come back to this case, which offers a nice interpretation of the resonance.
It corresponds approximately to a bound state of a chosen channel (hence justifying the discrete values of the resonant energies) which can decay into the open channels, an intuitive picture of a resonance, which corresponds very approximately to most of the dynamically generated states in the chiral unitary approach.

It is also interesting to obtain the relationship of the couplings and the wave functions at the origin.
For the case of bound states, it was particularly simple and we found\footnote{The reader will note that in Eq.~(118) of \cite{conenrique} there are two equations that should be separated by a space but they are not (the some happens in Eq.~(120)).}\cite{conenrique} $g_iG_{ii}^\alpha=\hat\psi_i$, where $\alpha$ refers to $G$ calculated for the energy of the bound state.
This relationship is tied to the discrete spectrum of the bound state and does not hold here.
Instead we found Eq.~(\ref{eq:39}), which we can rewrite as
\begin{eqnarray}
\hat\psi_i&=&(1-Gv)^{-1}_{i1}=[v^{-1}(v^{-1}-G)^{-1}]_{i1}\nonumber\\
&=&(v^{-1}t)_{i1}\equiv (v^{-1})_{ij}\frac{g_jg_1}{\sqrt{s}-M_{\rm R}+i\Gamma/2}~~.
\label{eq:g}
\end{eqnarray}
The last equation relates the couplings to the wave functions at the origin.
The value $\hat\psi_i$ of Eq.~(\ref{eq:g}) is tied to the choice of normalization made by us in Eq.~(\ref{eq:10}).
A more useful relationship, independent of the precise normalization, is given by
\begin{equation}
\frac{\hat\psi_i}{\hat\psi_k}=\frac{(v^{-1})_{ij}g_j}{(v^{-1})_{kj}g_j}~\to~\hat\psi_i=\alpha(v^{-1})_{ij}g_j
\end{equation}
which can be equivalently cast as
\begin{equation}
\frac{g_i}{g_k}=\frac{v_{ij}\hat\psi_j}{v_{kj}\hat\psi_j}~~.
\label{eq:ale}
\end{equation}
In the particular case where one channel $a$ is dominant over the others, $|v_{aa}|\gg|v_{ij}|$ ($i,~j\ne a$) and $|v_{ai}|\gg|v_{ij}|~(i,~j\ne a)$, Eq.~(\ref{eq:ale}) reads as
\begin{equation}
\frac{g_i}{g_a}=\frac{v_{ia}\hat\psi_a}{v_{aa}\hat\psi_a}=\frac{v_{ia}}{v_{aa}}~~.
\label{eq:eme}
\end{equation}

\subsection{Intuitive picture of a resonance as a bound state of a channel decaying into open ones}
Let us assume that we have just one bound channel, $a$, and we look for bound states of it.
We will solve the Schr$\ddot{\rm o}$dinger equation and find different discrete eigenenergies $E_\alpha$ for different eigenstates $\psi_\alpha$.
Let us just take one of these states and let us assume that $|v_{aa}|\gg |v_{ij}|$ ($i,~j\ne a$).
The picture is that we have essentially a bound state that decays into the channels $i\ne a$ (assume the rest of channels open).
An extreme picture of that would be an electron in an excited state of an atom which decays emitting a photon.
To a very large extend we can consider this state just a bound state of an electron, even if technically it should be considered a resonance because it couples to channels in the continuum at the states gets a width.

The width of the resonance can be calculated from Eq.~(\ref{eq:50}).
To simplify the formulation let us assume that we have the bound channel $a$ and the open channel 1 and neglect $v_{11}$ versus $v_{1a}$ and $v_{aa}$.
Eq.~(\ref{eq:t}) gives now:
\begin{eqnarray}
t&=&\frac{1}{1-v_{aa}G_a-v_{1a}^2G_1G_a}
\bordermatrix{
&&&\cr
&v_{1a}^2G_a&v_{1a}&\cr
&v_{1a}&v_{1a}^2G_1+v_{aa}&\cr
}\\
t_{aa}&\simeq&\frac{v_{aa}}{1-v_{aa}G_a-\partial G_a/\partial E\Bigr|_{E=E_{\rm R}}v_{aa}(E-E_{\rm R})-v_{1a}^2G_1G_a\Bigr|_{E=E_{\rm R}}}
\end{eqnarray}
where, since for the energy $E_{\rm R}$ we have approximately a bound state of $a$, according to \cite{conenrique} we will have $v_{aa}G_a=1$ and hence
\begin{equation}
t_{aa}=\frac{-1}{\partial G_a/\partial E}\frac{1}{E-E_{\rm R}-\frac{v_{1a}^2G_1G_a}{v_{aa}(-\partial G_a/\partial E)}\Bigr|_{E=E_{\rm R}}}\simeq \frac{g^2_a}{E-E_{\rm R}+i\Gamma_1/2}
\end{equation}
which leads to
\begin{equation}
g^2_{a}=\frac{1}{-\partial G_a/\partial E\Bigr|_{E=E_{\rm R}}}~;~\frac{\Gamma_1}{2}=\left.\frac{{\rm Im(}v_{1a}^2G_1G_a)}{v_{aa}\frac{\partial G_a}{\partial E}}\right|_{E=E_{\rm R}}~~.
\label{eq:86}
\end{equation}
Considering that from Eq.~(\ref{eq:G}) we find
\begin{equation}
{\rm Im~}G_1=-4\pi^2\mu_1k_1~~.
\end{equation}
From Eqs.~(\ref{eq:76}), (\ref{eq:ese}) and (\ref{eq:86}) we find
\begin{equation}
\frac{\Gamma_1}{2}=4\pi^2\mu_1k_1\frac{v_{1a}^2g_a^2G_a}{v_{aa}}=4\pi^2\mu_1k_1\frac{g_1^2v_{aa}^2G_a}{v_{aa}}=4\pi^2\mu_1k_1g_1^2
\label{eq:vale}
\end{equation}
where in the last step of Eq.~(\ref{eq:vale}) we have used Eq.~(\ref{eq:eme}) and $v_{aa}G_a=1$ for the bound state.
Eq.~(\ref{eq:vale}) is the same as Eq.~(\ref{eq:59}) upon establishing the equivalence of the couplings in Field Theory and Quantum Mechanics of Eq.~(\ref{eq:d}) for $i=1$,
\begin{equation}
\tilde g_1^2=32 \pi^3\mu_1\sqrt{s}(2M_{\rm B1})^{-1}g_1^2~~.
\end{equation}

Another way to proceed, which stresses the picture that we have, is a straightforward derivation of the probability to decay into the channel $j$ of the bound state $a$.
The standard procedure \cite{sakurai} starts from the $S$ matrix
\begin{eqnarray}
S&=&1-i\int^\infty_{-\infty}V_I dt+\cdots\nonumber\\
\langle j|S|a\rangle&=&-i\int^\infty_{-\infty}e^{-iE_jt}e^{iE_at}\langle j|V|a\rangle dt\nonumber\\
&=&-i2\pi\delta(E_j-E_a)\int d^3p\int d^3p'\langle j|\mbox{\boldmath $p$}\rangle \langle \mbox{\boldmath $p$}|V|\mbox {\boldmath $p'$}\rangle \langle\mbox{\boldmath $p'$}|a\rangle\nonumber\\
&=&-i2\pi\delta(E_j-E_a)v_{ja}\int_{p'<\Lambda}d^3p'\langle\mbox{\boldmath $p'$}|a\rangle=-i2\pi\delta(E_j-E_a)v_{ja}\hat\psi_a
\end{eqnarray}
where we have used $\int_{p<\Lambda} d^3p\langle j|\mbox{\boldmath $p$}\rangle=1$ for the asymptotic free $|j\rangle$ state ($|\mbox{\boldmath $p$}_j\rangle$) and must remember to sum over $|j\rangle$ in $|S|^2$ by means of $d^3p_j$ to be consistent with our normalization.
With the standard procedure to deal with $(\delta(E-E_j))^2$ we find
\begin{equation}
\Gamma=\int d^3p_j2\pi\delta(E_j-E_a)\frac{v_{ja}\hat\psi_av_{ja}\hat\psi_a^*}{\langle \psi_a|\psi_a\rangle}
\end{equation}
where we have divided by the norm squared of the wave function of the bound state $a$.
From \cite{conenrique} we know that for the bound state
\begin{equation}
\langle \psi_a|\psi_a\rangle=\int_{p<\Lambda}d^3p(\frac{1}{E_a-M_a-\mbox{\boldmath $p$}^2/2\mu_a})^2\sum_lv_{al}\hat\psi_l\sum_mv_{am}\hat\psi_m^*
\end{equation}
which, upon the approximations used that $|v_{aa}|\gg|v_{ij}|~(i,~j\ne a)$, reads as
\begin{equation}
\langle\psi_a|\psi_a\rangle=-\frac{\partial G_a}{\partial E}\Bigr|_{E=E_a}v_{aa}\hat\psi_av_{aa}\hat\psi_a^*
\end{equation}
and using the fact that $g_a^2=\Bigl[(-\partial G_a/\partial E)_{E=E_a}\Bigr]^{-1}$ we find
\begin{equation}
\Gamma=8\pi^2\mu_jp_j\frac{g^2_av_{ja}v_{ja}}{v_{aa}v_{aa}}=8\pi^2\mu_jp_jg_j^2
\end{equation}
where in the last step we used Eq.~(\ref{eq:eme}).
Thus, the result is the same as in Eq.~(\ref{eq:vale}).

The derivation has served to see that this result holds in the case that $a$ is a dominant channel with moderate decay into open ones, providing this intuitive picture for the resonant states in coupled channels.

\section{Application to the two $\Lambda(1405)$ states}

\subsection{The $\Lambda(1405)$ and $\Lambda(1670)$ resonances in the Chiral Unitary Model}
We apply the formalism explained in the former sections to study the wave functions of the two $\Lambda(1405)$ states generated in the chiral unitary approach \cite{carmina2,carmenjuan,cola}.
For the description of the $\Lambda(1405)$ states, we use the chiral unitary approach~\cite{oset,cola}.
In this model there are two $\Lambda(1405)$ states dynamically generated in the coupled channels of meson-baryon scattering,~${\bar K}N,~\pi\Sigma,~\eta\Lambda$ and $K\Xi$.
The most interesting thing in this model is that two poles exist around the $\Lambda(1405)$ energy region at $z_1=(1390,-i66)$ MeV and $z_2=(1426,-i16)$ MeV \cite{cola}.
These two poles are also found in all the works on the chiral unitary approach \cite{borasoy,ollerKa,borasoyulf} that followed Ref.~\cite{cola},
even when higher order terms in the chiral Lagrangians are considered.
Experimental support for these states has been shown in \cite{magas,sekihara}.
The electromagnetic mean squared radii of $\Lambda(1405)$ are calculated in Ref.~\cite{sekihara2} and are shown to be much larger than that of ground state baryons.
The $\Lambda(1670)$ was first reported as a dynamically generated resonance in Refs.~\cite{carmenjuan,oset} and has been corroborated in following works~\cite{carmina2,Hyodo:2006kg}.

\subsection{Wave function in coordinate space}
We study here the wave functions in $I=0$ of the ${\bar K}N,~\pi\Sigma,~\eta\Lambda$, $K\Xi$ channels.
Since the $\Lambda(1405)$ is observed in the $\pi\Sigma$ spectra in experiment, we consider that the $\pi\Sigma$ channel is the scattering state that we have called channel 1.

\begin{figure}[htpd]
\begin{center}
\includegraphics[width=10.0cm,height=15.0cm]{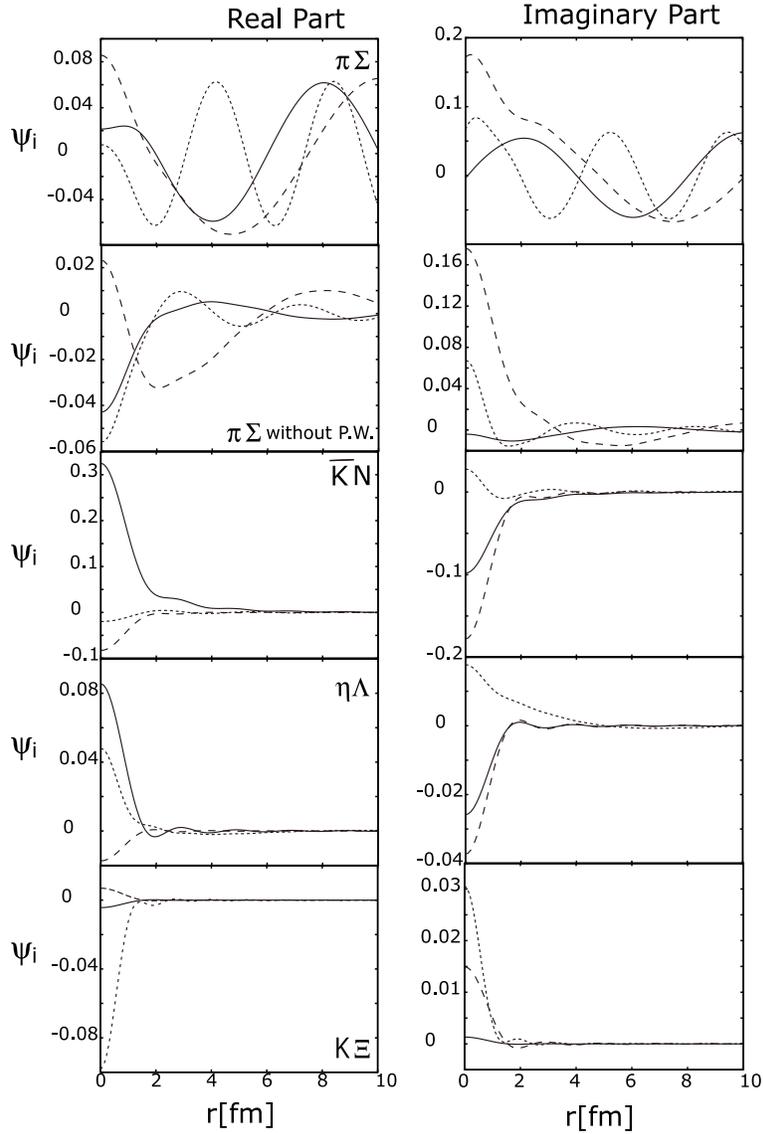}
\caption{\label{fig:WF}Wave functions in coordinate space. 
Solid lines and dashed lines show the results obtained at $E=1426$ MeV and at $E=1390$ MeV for the $\Lambda(1405)$ resonances, and dotted lines show that at $E=1680$ MeV for the $\Lambda(1670)$ state.
The figures of the first line contain the full wave function of $\pi\Sigma$ (initial plus scattered), while the second line shows only the scattered wave of the $\pi\Sigma$ state}
\end{center}
\end{figure}

We show in Fig.~\ref{fig:WF} the wave function in coordinate space at the pole energies of the two $\Lambda(1405)$, together with that of the $\Lambda(1670)$.
From Fig.~\ref{fig:WF}, we find that the ${\bar K}N$ components dominate at 1426 MeV while the $\pi\Sigma$ components are dominant at 1390 MeV.
This is consistent with the findings of \cite{carmina2,carmenjuan,cola, borasoy, ollerKa,borasoyulf} that the pole at higher energies couples most strongly to ${\bar K}N$ while the one at lower energies couples mostly to $\pi\Sigma$.

The curves correspond to these different energies, which are the energies where we find the poles of the $\Lambda(1390)$, $\Lambda(1426)$ and $\Lambda(1670)$.
We can see how the wave function concentrates close to the origin, and both for bound channels, as well as for open channels, fades away rapidly beyond 2 $fm$, providing a spatial picture of the distribution of the particles of the different channels building up the resonances.

In the case of the $\Lambda(1670)$ we can see in the last line of Fig.~\ref{fig:WF} that the dominant component is the $K\Xi$ bound state.
This resonance would very approximately qualify as a $K\Xi$ bound state, as also suggested in Ref.~\cite{oset} based on the large couplings of the resonance to that state.

\section{Response function and form factors}
\begin{figure}[htpd]
\begin{center}
\includegraphics[width=5.0cm,height=6.0cm]{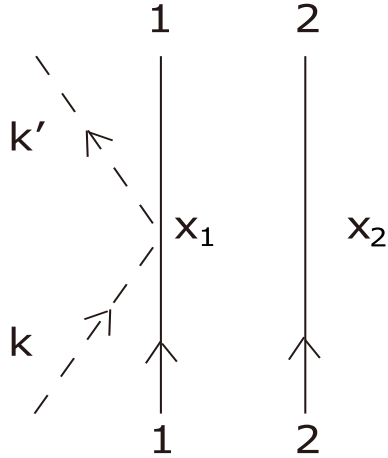}
\caption{\label{fig:new}
Coupling on external source to a molecular component}
\end{center}
\end{figure}

As a further application of the formalism, let us compute the response function of one resonance state to an external scalar source.
Let us take one of the components (one channel), the final expression will contain the sum of the partial response functions weighted by its coupling to the resonance squared, as we shall see.
Let us assume the channel to be a bound state for the moment, and let us couple the external scalar source, with strength unity, to one of the particles, particle 1 in Fig.~\ref{fig:new}.
The $S$ matrix for this diagram is given by 
\begin{equation}
S=\int d^4x_1\frac{1}{\sqrt{2\omega_{p_1}}}e^{-ip^0_1x^0_1}\phi_1(\mbox{\boldmath $x$}_1)\frac{1}{\sqrt{2\omega_{p'_1}}}e^{ip'^{0}_1x^0_1}\phi_1(\mbox{\boldmath $x$}_1)\frac{1}{\sqrt{2\omega_k\cal{V}}}e^{-ikx_1}\frac{1}{\sqrt{2\omega _{k'}\cal{V}}}e^{ik'x_1}(-it_1),
\label{eq:luis19}
\end{equation}
where $\cal{V}$ is the volume of a box where we normalize to unity our wave functions.
In Eq.~(\ref{eq:luis19}) $t_1\equiv 1$ for our scalar source.
We can multiply by
\begin{equation}
\int d^3x_2 \phi_2(\mbox{\boldmath $x$}_2)\phi_2(\mbox{\boldmath $x$}_2)=1
\label{eq:luis20}
\end{equation}
and perform the $x^0_1$ integration which provides the $\delta$ of conservation of energy.
Furthermore we can now write
\begin{equation} 
\phi_1(\mbox{\boldmath $x$}_1)\phi_2(\mbox{\boldmath $x$}_2)=\frac{1}{\sqrt{\cal{V}}}e^{i \mbox{\boldmath \scriptsize$K$}_m\cdot\mbox{\boldmath \scriptsize$R$}}\phi(\mbox{\boldmath $x$})
\end{equation}
where $\mbox{\boldmath $R$},~\mbox{\boldmath $x$}$ are the CM and relative coordinates given by
\begin{eqnarray}
\mbox{\boldmath $R$}&=&\frac{m_1\mbox{\boldmath $x$}_1+m_2\mbox{\boldmath $x$}_2}{m_1+m_2}\\
\mbox{\boldmath $x$}&=&\mbox{\boldmath $x$}_2-\mbox{\boldmath $x$}_1
\end{eqnarray}
and $\mbox{\boldmath $K$}_m,~\phi$ are the total momentum of the molecule and its relative wave function.

After performing the $d^3x_1d^3x_2\to d^3Rd^3x$ integrations we obtain
\begin{equation}
S=-it_1\frac{1}{{\cal V}^2}\frac{1}{\sqrt{2\omega_{p_1}}}\frac{1}{\sqrt{2\omega_{p'_1}}}\frac{1}{\sqrt{2\omega_k}}\frac{1}{\sqrt{2\omega_{k'}}}(2\pi)^4\delta^4(k+K_m-k'-K_m')F\Bigl(\frac{m_2}{m_1+m_2}(\mbox{\boldmath $k$}-\mbox{\boldmath $k$}')\Bigr)
\label{eq:luis26}
\end{equation}
where $F(\mbox{\boldmath $q$})$ is the form factor given by
\begin{equation}
F(\mbox{\boldmath $q$})=\int d^3x \phi^2(\mbox{\boldmath $x$})e^{-i \mbox{\boldmath \scriptsize$q$}\cdot\mbox{\boldmath \scriptsize$x$}}
\label{eq:FF_a}
\end{equation}
assuming the wave function real for the bound state, and $K_m^0-K_m^{'0}=p_1^0-p'^{0}_1$.
The scalar source can couple to particle 2 and we would have to sum this contribution, which is trivially obtained by exchanging the 1 and 2 indices in the former expressions.

It is useful to go to momentum space to evaluate Eq.~(\ref{eq:FF_a}). We have
\begin{eqnarray}
\phi(\mbox{\boldmath $x$})&=&\langle \mbox{\boldmath $x$}|\phi\rangle=\int\frac{d^3p}{(2\pi)^{3/2}}e^{i \mbox{\boldmath \scriptsize$p$}\cdot\mbox{\boldmath \scriptsize$x$}}\langle \mbox{\boldmath $p$}|\psi\rangle\\
\langle\mbox{\boldmath $p$}|\psi\rangle&=&v\frac{\theta(\Lambda-p)}{E-\omega_1(p)-\omega_2(p)}\int_{k<\Lambda}d^3k\langle\mbox{\boldmath $k$}|\psi\rangle~~.
\end{eqnarray}
Hence, although we can take $\phi(\mbox{\boldmath $x$})$ real for a bound components (we will see latter on the generalization to open channels), formally we can write
\begin{eqnarray}
F(\mbox{\boldmath $q$})&=&\int d^3x\phi(\mbox{\boldmath $x$})\phi^*(\mbox{\boldmath $x$})e^{-i \mbox{\boldmath \scriptsize$q$}\cdot\mbox{\boldmath \scriptsize$x$}}\nonumber\\
&=&\int d^3x\int\frac{d^3p}{(2\pi)^{3/2}}e^{i \mbox{\boldmath \scriptsize$p$}\cdot\mbox{\boldmath \scriptsize$x$}}\frac{\theta(\Lambda-p)}{E-\omega_1(p)-\omega_2(p)}\nonumber\\
&&\times\int\frac{d^3p'}{(2\pi)^{3/2}}e^{-i \mbox{\boldmath \scriptsize$p'$}\cdot\mbox{\boldmath \scriptsize$x$}}\frac{\theta(\Lambda-p')}{E-\omega_1(p')-\omega_2(p')}e^{-i \mbox{\boldmath \scriptsize$q$}\cdot\mbox{\boldmath \scriptsize$x$}}\nonumber\\
&=&\int d^3p \frac{\theta(\Lambda-p)\theta(\Lambda-|\mbox{\boldmath $p$}-\mbox{\boldmath $q$}|)}{(E-\omega_1(p)-\omega_2(p))(E-\omega_1(\mbox{\boldmath $p$}-\mbox{\boldmath $q$})-\omega_2(\mbox{\boldmath $p$}-\mbox{\boldmath $q$}))}
\label{eq:FF_d}
\end{eqnarray}
up to a normalization, easily restored demanding that $F(q=0)=1$.

\begin{figure}[htpd]
\begin{center}
\includegraphics[width=6.0cm,height=5.0cm]{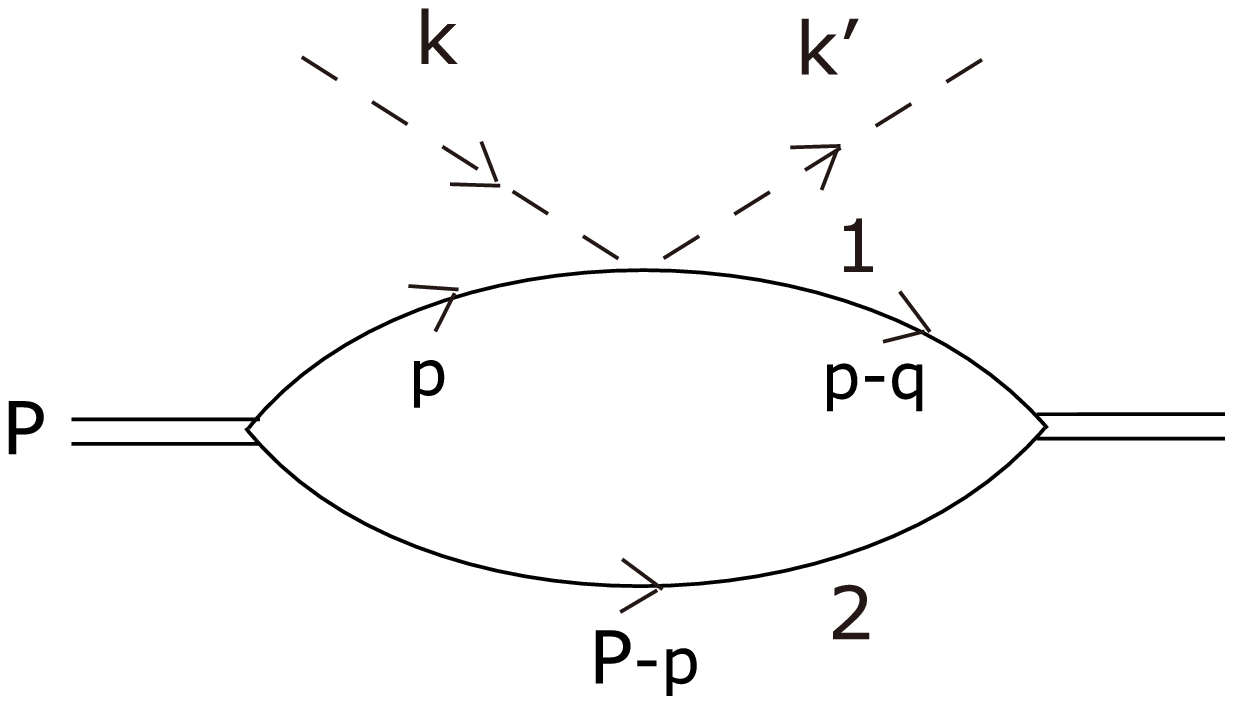}
\caption{\label{fig:newnew}}
\end{center}
\end{figure}

Let us now compare this result with what one would obtain in a field theoretical approach in which the scalar source couples to the components of the resonance.
Diagrammatically the process is depicted in Fig.~\ref{fig:newnew}.
The response function from interaction with particle 1 will be (taking $\mbox{\boldmath $P$}=0$, the total three momentum of the resonance)
\begin{equation}
-i\int\frac{d^4p}{(2\pi)^4}\frac{\theta(\Lambda-p)}{p^0-\omega_1(p)+i\epsilon}\frac{\theta(\Lambda-|\mbox{\boldmath $p$}-\frac{m_2}{m_1+m_2}\mbox{\boldmath $q$}|)}{p^0-q^0-\omega_1(\mbox{\boldmath $p$}-\mbox{\boldmath $q$})+i\epsilon}\frac{1}{P^0-p^0-\omega_2(p)+i\epsilon}
\label{eq:FF_107}
\end{equation}
which, upon, contour integration of the $p^0$ variable on the upper half circle gives
\begin{equation}
\int\frac{d^3p}{(2\pi)^3}\frac{\theta(\Lambda-p)}{P^0-\omega_2(p)-\omega_1(p)+i\epsilon}\frac{\theta(\Lambda-|\mbox{\boldmath $p$}-\frac{m_2}{m_1+m_2}\mbox{\boldmath $q$}|)}{P^0-q^0-\omega_1(\mbox{\boldmath $p$}-\mbox{\boldmath $q$})-\omega_2(p)+i\epsilon}
\label{eq:FF_c}
\end{equation}
with $q^0$ the excitation energy carried by the external source in this reference frame, such that $P^0-q^0=E_R(\mbox{\boldmath $q$})$, $q^0=M_R-E_R(\mbox{\boldmath $q$})=-\mbox{\boldmath $q$}^2/2(m_1+m_2)$.

Even if Eq.~(\ref{eq:FF_d}) and Eq.~(\ref{eq:FF_c}) do not look the same, one can see that in the nonrelativistic limit one has
\begin{eqnarray}
&&\mbox{\boldmath $p$}_1=m_1\mbox{\boldmath ${\dot x}$}_1;~\mbox{\boldmath $p$}_2=m_2\mbox{\boldmath ${\dot x}$}_2;~\mbox{\boldmath $P$}=M\mbox{\boldmath ${\dot R}$}=\mbox{\boldmath $p$}_1+\mbox{\boldmath $p$}_2\nonumber\\
&&\mu=\frac{m_1m_2}{m_1+m_2},~\mbox{\boldmath $p$}=\mu\mbox{\boldmath ${\dot x}$}=\frac{m_1}{M}\mbox{\boldmath $p$}_2-\frac{m_2}{M}\mbox{\boldmath $p$}_1~~,
\label{eq:hola}
\end{eqnarray}
and then the field theoretical approach with the loop function simply provides the form factor that one obtains in the wave function approach.
The argument of the second $\theta$ function in Eqs.~(\ref{eq:FF_107}) and (\ref{eq:FF_c}) corresponds to the relative momentum for particles 1 and 2, having momenta $\mbox{\boldmath $p$}-\mbox{\boldmath $q$}$ and $-\mbox{\boldmath $p$}$, as shown as Fig.~\ref{fig:newnew}, which is $-\mbox{\boldmath $p$}+\frac{m_2}{m_1+m_2}\mbox{\boldmath $q$}$.

One subtlety is worth mentioning at this point.
In the case of open channels in the wave function method, the evaluation of the form factor with the used $\int\phi(\mbox{\boldmath $x$})\phi^*(\mbox{\boldmath $x$})\exp(-i\mbox{\boldmath $q$}\cdot\mbox{\boldmath $x$})d^3x$ would lead to the expression of Eq.~(\ref{eq:FF_c}) with a $-i\epsilon$ instead of $i\epsilon$ in the second factor.
The field theoretical approach, which provides an appropriate formalism for these processes, keeps the $+i\epsilon$ in the two propagators.
This means that in the case of interaction of an external source with the open channels of a resonance the response function does not involve the ordinary form factor involving $\phi\phi^*$ but something else, which in the case of bound channels is the ordinary form factor.
The situation can also be interpreted from the quantum mechanical side, since in the case of decay into an open channel, the $\phi^*$ conjugate of the wave function of the final state is an outgoing solution of the Schr$\ddot{\rm o}$dinger equation with the potential involved (complex to account for inelastic channels) and not of the complex conjugate of the Schr$\ddot{\rm o}$dinger equation~\cite{Straub:1992yw}.
In the case that we have several channels building up a resonance, the response function is given by
\begin{equation}
R(\mbox{\boldmath $q$})=\sum_{\rm channels}g_i\tilde{G_i}(E,\mbox{\boldmath $q$})g_i
\end{equation}
with $g_i$ the coupling of the resonance to each channel and 
\begin{equation}
\tilde{G_i}=\int \frac{d^3p}{(2\pi)^{3/2}}\frac{\theta(\Lambda-p)\theta(\Lambda-|\mbox{\boldmath $p$}-\frac{m_{2i}}{m_{1i}+m_{2i}}\mbox{\boldmath $q$})|)}{(E-\omega_1(p)-\omega_2(p)+i\epsilon)(\sqrt{E^2+\mbox{\boldmath $q$}^2}-\omega_1(\mbox{\boldmath $p$}-\mbox{\boldmath $q$})-\omega_2(p)+i\epsilon)}
\end{equation}
In the response function we would have to sum over the second particle too.

\begin{figure}[htpd]
\begin{center}
\includegraphics[width=12.0cm,height=9.0cm]{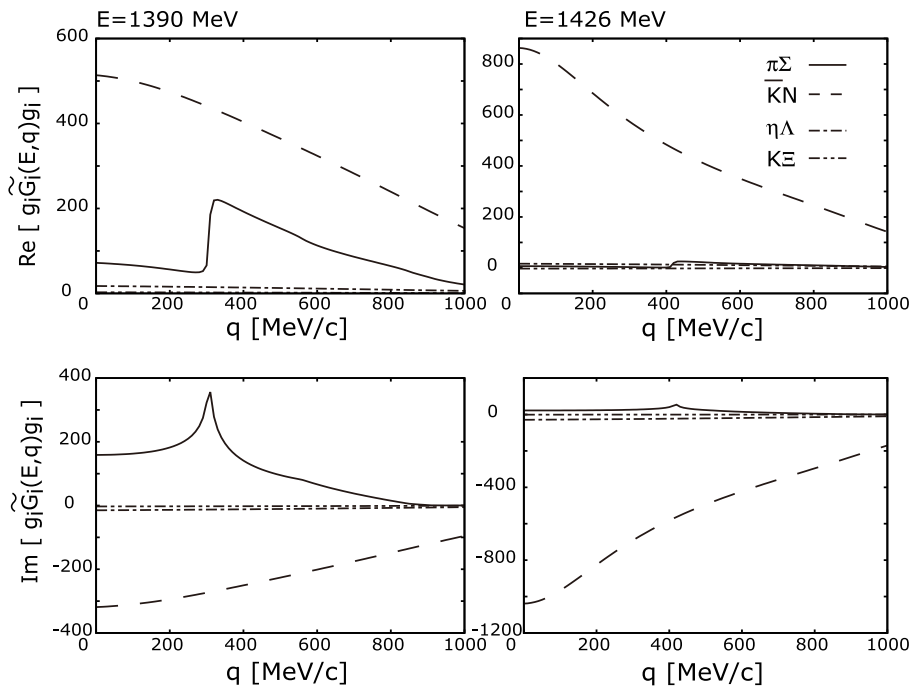}
\caption{\label{fig:meson}
Contribution of the meson part of each channel to the response function for the two $\Lambda(1405)$ states.}
\end{center}
\end{figure}

\begin{figure}[htpd]
\begin{center}
\includegraphics[width=12.0cm,height=9.0cm]{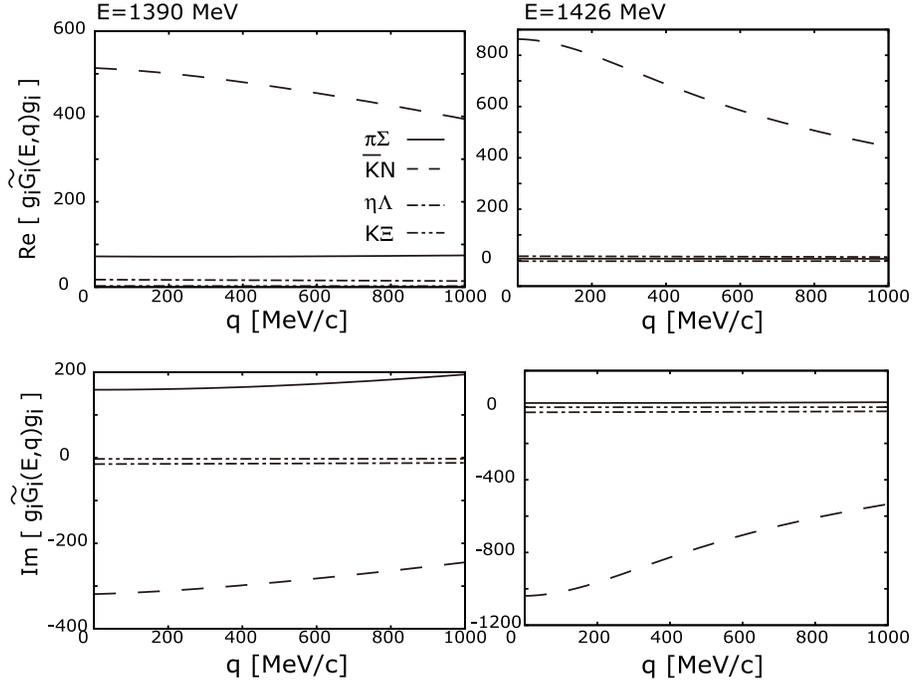}
\caption{\label{fig:baryon}
Contribution of the baryon part of each channel to the response function for the two $\Lambda(1405)$ states.}
\end{center}
\end{figure}

\begin{figure}[htpd]
\begin{center}
\includegraphics[width=12.0cm,height=9.0cm]{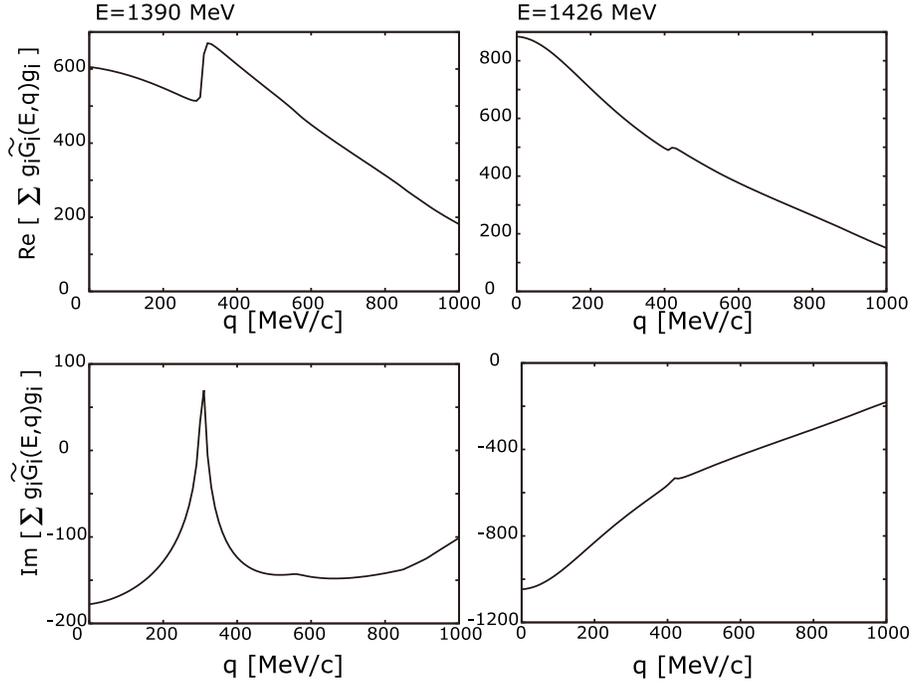}
\caption{\label{fig:meson_tot}
Sum of all contribution from the meson part of the two resonances.}
\end{center}
\end{figure}

\begin{figure}[htpd]
\begin{center}
\includegraphics[width=12.0cm,height=9.0cm]{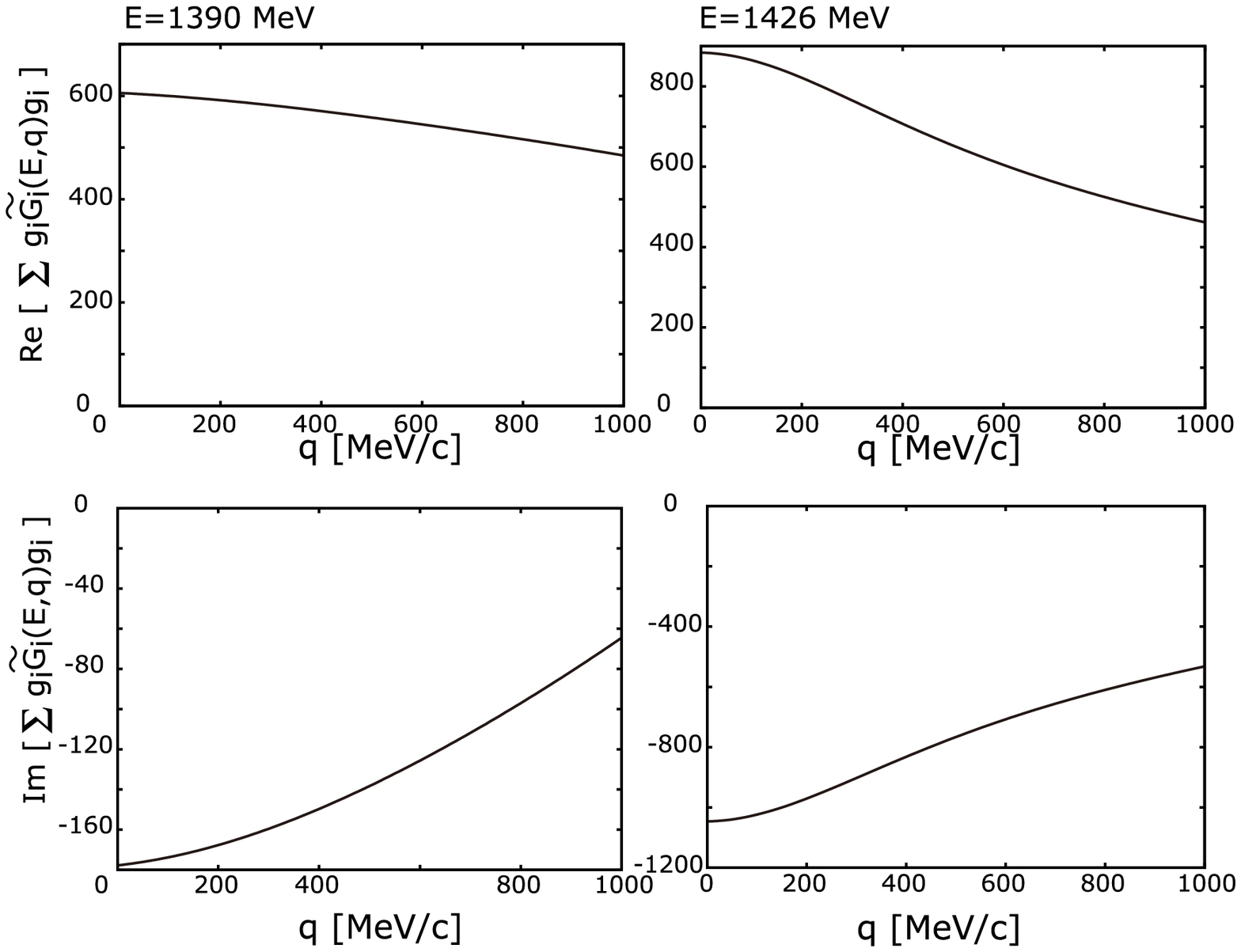}
\caption{\label{fig:baryon_tot}
Sum of all contribution from the baryon part of the two resonances.}
\end{center}
\end{figure}

We show results in Figs.~\ref{fig:meson} to \ref{fig:baryon_tot}.

In Fig.~\ref{fig:meson} we show the contribution to the response function from the meson component of each channel for two different energies, which correspond to the two $\Lambda(1405)$ resonances of our approach.
We see a gradual fall down of the response function with $q$, approaching zero at momentum of the order of 1 GeV.
In Fig.~\ref{fig:baryon} we show the corresponds contributions from the baryon part of the resonances.
Finally in Figs.~\ref{fig:meson_tot},~\ref{fig:baryon_tot} we show the sum of the contributions of the different channels from the meson and baryon parts respectively.
One observes differences in the behavior of the scalar response function for the two resonances, with a faster fall down for the $\Lambda(1420)$ case.
Although the scalar form factor is different than the charge form factor, it is interesting to note that there are similarities of our results with the charge form factors evaluated in \cite{sekihara2}, where the one of the $\Lambda(1420)$ has also a faster fall down with $q$ than the one of the $\Lambda(1390)$.

\section{Summary}
    In this paper we have developed a formalism to deal with coupled channels in a unitary approach by paying a special attention to constructing the wave functions in the different channels in the case that there are resonances dynamically generated. The paper generalizes what was found before for only bound coupled channels. Here we have bound and open channels and the formalism is subtly different, since contrary to the case of bound states, where only discrete energies are allowed, here we have a continuous energy variable. Many of the results obtained for bound states do not hold for the resonance states. One of the things we do is to identify the meaning of a resonance in the coupled channel approach, and it emerges as an approximate bound state of a coupled channel which can decay into the open ones.  The formalism developed is easy, practical and useful.  A separable potential in coordinate space is chosen which leads to an on shell factorization of the Bethe Salpeter equations (Lippmann Schwinger in the nonrelativistic form), which allows to convert the coupled channel integral equations into trivial algebraic equations. The wave functions in momentum space are then found as trivial analytic functions, from where the wave functions in coordinate space can be easily evaluated.  The couplings of the resonance to the different channels are related to the wave function at the origin and interesting relationships between these couplings are obtained.
     We also study the issue of couplings of the resonances to states outside the space of the building channels and justify results used before in the Literature, setting the limits for their application. Similarly, we also face the issue of final state interaction within the coupled channel formalism and find again a justification for results used in the Literature, setting again the limits of applicability. 
     
     As an application of the formalism, we tackle the problem of the two $\Lambda(1405)$ and the $\Lambda(1670)$ states dynamically generated in the chiral unitary approach from  the $\pi \Sigma$, $\bar{K} N$, $\eta \Lambda$, and $K \Xi$  interaction. We evaluate the wave functions in coordinate space for the first time, giving an intuitive idea of the wave functions and the spatial distribution of the particles of the different channels.  
We have also evaluated the response function of the resonances to an external scalar source, together with the form factors involved.
     
     We envisage practical applications of the formalism in any problem where one has to deal with the interaction of external sources with the dynamically generated resonances. In particular in the building up of multihadron states using iteratively the Fixed Center Approximation to the Faddeev equations, where other methods would turn out technically prohibitive. The resulting wave functions can also be used to evaluate static properties of the resonances and different form factors, etc, which might be studied experimentally in the future.

\section{Acknowledgments}
  This work is partly supported 
by DGICYT Contracts Nos. FIS2006-03438 and FEDER funds, FIS2008-01143 and CSD2007-00042 (CPAN), the Generalitat Valenciana in the program Prometeo 
and the EU Integrated
Infrastructure Initiative Hadron Physics Project under contract
RII3-CT-2004-506078.
This research is part of the European
 Community-Research Infrastructure Integrating Activity ``Study of
 Strongly Interacting Matter'' (acronym HadronPhysics2, Grant
 Agreement n. 227431) 
 under the Seventh Framework Programme of EU.  

\end{document}